\documentclass{article}
\usepackage{spconf,amsmath,graphicx}
\usepackage{amsfonts}
\usepackage{booktabs}
\usepackage{color}
\usepackage{caption}
\usepackage{subcaption}
\usepackage{multirow}
\usepackage[toc,page]{appendix}
\usepackage{comment}
\usepackage{url}

\title{Similarity Analysis of Self-Supervised Speech Representations}
\name{Yu-An Chung$^{1}$, Yonatan Belinkov$^{2*}$\thanks{* Supported by the Viterbi Fellowship in the Center for Computer Engineering at the Technion.}, James Glass$^{1}$}
\address{
  $^{1}$MIT Computer Science and Artificial Intelligence Laboratory, Cambridge, MA 02139, USA\\
  $^{2}$Technion Henry and Marilyn Taub Faculty of Computer Science, Haifa 3200003, Israel\\
  {\small \texttt{\{andyyuan,glass\}@mit.edu, belinkov@technion.ac.il}}
}
\begin{document}
\ninept
\maketitle
\begin{abstract}
Self-supervised speech representation learning has recently been a prosperous research topic.
Many algorithms have been proposed for learning useful representations from large-scale unlabeled data, and their applications to a wide range of speech tasks have also been investigated.
However, there has been little research focusing on understanding the properties of existing approaches.
In this work, we aim to provide a comparative study of some of the most representative self-supervised algorithms.
Specifically, we quantify the similarities between different self-supervised representations using existing similarity measures.
We also design probing tasks to study the correlation between the models' pre-training loss and the amount of specific speech information contained in their learned representations.
In addition to showing how various self-supervised models behave differently given the same input, our study also finds that the training objective has a higher impact on representation similarity than architectural choices such as building blocks~(RNN/Transformer/CNN) and directionality~(uni/bidirectional).
Our results also suggest that there exists a strong correlation between pre-training loss and downstream performance for some self-supervised algorithms.
\end{abstract}
\begin{keywords}
Self-supervised learning, speech representation learning, unsupervised pre-training, comparative analysis
\end{keywords}
\section{Introduction}
Self-supervised learning is a form of unsupervised learning that treats the input or modifications of the input as learning targets.
Thanks to this property, self-supervised learning can leverage large-scale unlabeled data for training, and has enjoyed success in learning high-level representations of data from different modalities~\cite{chen2020simple,devlin2019bert,baevski2020wav2vec}.

Recently, self-supervised approaches for learning speech representations have received great research attention.
Methods like contrastive predictive coding~\cite{oord2018representation}, autoregressive predictive coding~\cite{chung2019unsupervised}, masked predictive coding~\cite{liu2020mockingjay,wang2020unsupervised,jiang2019improving}, and problem-agnostic speech encoder~\cite{pascual2019learning} have been shown to be capable of learning representations that capture high-level properties of speech that are not easily accessible from surface features such as audio waveforms and spectrograms.
These methods have been further extended or improved for tackling a wide range of speech applications, including speech recognition~\cite{baevski2020effectiveness,chung2020generative,ling2020deep,jiang2020further,song2020speech}, speech translation~\cite{nguyen2020investigating,wu2020self}, speaker verification~\cite{ravi2020exploring}, unsupervised unit discovery~\cite{feng2020unsupervised}, and unsupervised phoneme segmentation~\cite{kreuk2020self}, to name a few.

Despite the recent progress in self-supervised speech representation learning, most of the effort is made to develop new algorithms or adapt existing methods to particular tasks, and only a few studies focus on reviewing existing approaches.
In this work, we aim to provide a comparative study on some of the most representative self-supervised algorithms: contrastive predictive coding~(CPC), autoregressive predictive coding~(APC), and masked predictive coding~(MPC).
Our analysis focuses on the following two aspects.
First, we hope to understand the similarity of different self-supervised representations.
To carry out this study, we adopt two similarity measures for quantifying the similarity of two given representations.
Although such a similarity analysis approach cannot discern absolute facts about the representations, it allows us to compare representations without subscribing to any specific type of information, and helps us answer questions like: Given the same input, how similar are different self-supervised representations? Which modeling choices, e.g., building blocks~(RNN/Transformer/CNN) and directionality~(uni/bidirectional), have a higher impact on similarity? How much does a model change when it is trained on more data?

Our second area of investigation examines, for each self-supervised algorithm, how well its pre-training loss correlates with downstream performance.
We use phonetic and speaker classification as probing tasks to measure the amount of phonetic and speaker information contained in the representations as a function of pre-training loss.
This study could be useful for model selection if there exists a strong correlation between them.

Only a few studies have focused on analyzing self-supervised models.
Chung et al.~(2020) proposed to incorporate vector quantization layers to restrict model capacity during pre-training so as to uncover a model's preference in preserving speech information for achieving a maximal self-supervised objective~\cite{chung2020vector}.
Bland{\'o}n and R{\"a}s{\"a}nen~(2020) studied the correlation between the self-supervised loss of APC and CPC and their performance on a phoneme discrimination task~\cite{blandon2020analysis}, which has the same goal as our second study.
However, neither of these two works investigated the similarity between different self-supervised representations.
For the correlation study, we also consider more self-supervised models with diverse modeling choices as compared to previous work~\cite{blandon2020analysis}.

Our analysis yields the following insights:
\begin{itemize}
  \item The objective has a higher impact on representation similarity than model architecture.
  \item Under the same objective, a model's directionality\\(uni/bidirectional) affects representation similarity more than its building blocks~(RNN/Transformer/CNN).
  \item Both APC and MPC both have a stronger correlations between pre-training loss and  phonetic and speaker classification performance than does CPC.
  \item While all models benefit from increasing the size of unlabeled training data, CPC is found to make use of these additional data more efficiently than APC and MPC.
\end{itemize}


\section{Analysis Methods}
\label{sec:analysis}
We study two aspects of self-supervised speech representation learning: (1) the similarity between representations learned by various models, and (2) how well their self-supervised pre-training loss correlates with downstream performance.
We describe our methods for analyzing these two aspects in Sections~\ref{sec:sim} and~\ref{sec:probe}, respectively.

\subsection{Approaches for measuring representation similarity}
\label{sec:sim}
Consider a pre-trained self-supervised model~$M$.
For an acoustic feature sequence~(in our case, a log Mel spectrogram)~$\mathbf{x} = (\mathbf{x}_{1}, \mathbf{x}_{2}, ..., \mathbf{x}_{T})$, where~$\mathbf{x}_{t}\in \mathbb{R}^{80}$, from a dataset~$D$, the model~$M$ transforms~$\mathbf{x}$ into a representation~$M(\mathbf{x}) = (\mathbf{m}_{1}, \mathbf{m}_{2}, ..., \mathbf{m}_{T})$, where~$\mathbf{m}_{t}\in \mathbb{R}^{512}$.
Given two representations extracted by two self-supervised models~$M^{(1)}$ and~$M^{(2)}$, a similarity measure outputs~$\mathrm{sim}(M^{(1)}(\mathbf{x}), M^{(2)}(\mathbf{x}))\in \mathbb{R}$ that quantifies their similarity.
Note that this approach does not require~$D$ to be annotated.

Existing similarity measures are proposed to capture different similarity notions.
Some focus on capturing the localization of information of two representations, which is usually done by comparing the behaviors of two individual elements~$\mathbf{m}^{(1)}_{i}$ and~$\mathbf{m}^{(2)}_{j}$ from~$M^{(1)}(\mathbf{x})$ and~$M^{(2)}(\mathbf{x})$, respectively~\cite{bau2019identifying}.
Other measures emphasize distributivity of information and find correlations between two representations~$M^{(1)}(\mathbf{x})$ and~$M^{(2)}(\mathbf{x})$ directly~\cite{kriegeskorte2008representational,kornblith2019similarity,raghu2017svcca,andrew2013deep}: if two representations behave similarly over all of their elements, their similarity will be high even if no two individual elements have similar behaviors.
In this work we focus on the latter case and adopt linear centered kernel alignment~({\tt \small lincka}; \cite{kornblith2019similarity}) and singular vector canonical correlation analysis~({\tt \small svcca};  \cite{raghu2017svcca}) as our similarity measures.
We choose these two since they are found to be comparable or better than other measures in prior studies for analyzing contextual word representation models~\cite{wu2020similarity}.

\subsection{Probing tasks for measuring phonetic and speaker content}
\label{sec:probe}
We consider phonetic and speaker classification for measuring the amount of phonetic and speaker content contained in a representation \cite{belinkov2017analyzing,wang2017does}.
Given a self-supervised model~$M$ pre-trained on an unlabeled dataset~$D_{1}$, we use~$M$ to extract features~$M(\mathbf{x}) = (\mathbf{m}_{1}, \mathbf{m}_{2}, ..., \mathbf{m}_{T})$, where~$\mathbf{m}_{t}\in \mathbb{R}^{512}$ for another dataset~$D_{2}$, and train a linear classifier using the extracted features as input.

For phonetic classification, the goal is to predict the phone identity of each frame in an input utterance.
For speaker classification, the extracted features are first averaged before being fed to the classifier, and the goal is to predict the speaker identity of the utterance.
The frame-level phone error rate and utterance-level speaker error rate on the test set of~$D_{2}$ indicate the amount of phonetic and speaker content contained in the representation, respectively.

\section{Experimental Setup}
\label{sec:setup}

\begin{table}[htbp]
  \caption{Information about various implementations of APC, MPC, and CPC to be compared in this work. All RNN and Transformer models have a hidden size of~512~(256 for forward and~256 for backward if bidirectional). For CPC, {\tt \small cpc-mixed\_spk-rnn} draws negative samples across speakers, while {\tt \small cpc-within\_spk-rnn} and {\tt \small cpc-within\_spk-cnn} draw negative samples from the same utterance as the target future frame. All building blocks have~3 layers unless otherwise stated. TRF stands for Transformer.}
  \label{tab:model_notations}
  \centering
  \resizebox{\columnwidth}{!}{
  \begin{tabular}{lccc}
    \toprule
    Model notation                    &  Objective  &  Building block  &  Uni- or Bi-dir \\
    \midrule
    {\tt \small apc-fw-rnn}           &  APC        &  GRU             &  Uni-dir \\
    {\tt \small apc-fw+bw-rnn}        &  APC        &  GRU             &  Bi-dir  \\
    {\tt \small apc-fw-trf}           &  APC        &  TRF decoder     &  Uni-dir \\
    {\tt \small apc-fw+bw-trf}        &  APC        &  TRF decoder     &  Bi-dir  \\
    {\tt \small mpc-birnn}            &  MPC        &  GRU             &  Bi-dir  \\
    {\tt \small mpc-trf}              &  MPC        &  TRF encoder     &  Bi-dir  \\
    {\tt \small cpc-mixed\_spk-rnn}   &  CPC        &  GRU             &  Uni-dir \\
    {\tt \small cpc-within\_spk-rnn}  &  CPC        &  GRU             &  Uni-dir \\
    {\tt \small cpc-within\_spk-cnn}  &  CPC        &  Same as~\cite{schneider2019wav2vec}  &  - \\
    \bottomrule
  \end{tabular}
  }
\end{table}

\subsection{Self-supervised models}
We consider some of the most representative models, including contrastive predictive coding~(CPC)~\cite{oord2018representation}, autoregressive predictive coding~(APC)~\cite{chung2019unsupervised}, and masked predictive coding~(MPC)~\cite{liu2020mockingjay,wang2020unsupervised,jiang2019improving}.

While there are additional models that have successfully been applied to speech applications, most of them are more or less an extension of the above models.
For example, Rivi{\`e}re et al.~(2020) improved CPC by modifying its batch normalization mechanism and replacing the linear prediction head with a~1-layer Transformer network~\cite{riviere2020unsupervised}.
Kawakami et al.~(2020) made CPC bidirectional~\cite{kawakami2020unsupervised}.
wav2vec~\cite{schneider2019wav2vec} is CPC with a convolutional architecture.
DeCoAR~\cite{ling2020deep} could be viewed as a bidirectional version of APC.
Chung and Glass~(2020) proposed an auxiliary loss serving as a regularizer to help APC generalize better~\cite{chung2020improved}.
Liu et al.~(2020) applied SpecAugment~\cite{park2019specaugment} to improve MPC's masking techniques~\cite{liu2020tera}.
Jiang et al.~(2020) combined APC and MPC to form a unified pre-training objective~\cite{jiang2020further}.
We leave the explorations of these extensions for future work.
Below we briefly review CPC, APC, and MPC.

\newcommand{\myparagraph}[1]{\vspace{.4em} \noindent \textbf{#1}\ }
\myparagraph{CPC \& APC}
Contrastive predictive coding~(CPC) and autoregressive predictive coding~(APC) share a similar methodology as both use an autoregressive model to learn representations through conditioning on the past context to make predictions of future information.
Their main difference lies in the manner in which they optimize the autoregressive model: while APC attempts to predict a future frame via L1 regression, CPC incorporates a proposal distribution for drawing negative samples, and learns representations containing information that most discriminates the future frame from the negative samples using a loss based on noise-contrastive estimation~\cite{gutmann2010noise}.
We mainly follow the original papers~\cite{oord2018representation,chung2019unsupervised} for implementing the models with small modifications described in~\cite{chung2019unsupervised}.

Since the objectives of APC and CPC are based on the notion of future prediction, bidirectional architectures are not applicable.
A simple method for making these models have access to context from both directions is to separately train a forward and backward APC/CPC model and concatenate their output representations as the final representations~(similar to how ELMo~\cite{peters2018deep} is trained for learning contextualized word embeddings).
This method has been explored for APC and CPC in~\cite{ling2020deep} and~\cite{kawakami2020unsupervised}, respectively.

\myparagraph{MPC}
Inspired by the masked language modeling technique from BERT~\cite{devlin2019bert}, masked predictive coding~(MPC) directly trains a bidirectional architecture by first masking parts of the input signals and then predicting them through conditioning on context from both directions.
Similar to APC, MPC is optimized by minimizing the frame-wise L1 distance between the predicted output and the original input before masking.
Transformer encoder~\cite{liu2020mockingjay,jiang2019improving} and bidirectional RNN~\cite{wang2020unsupervised} have both been used to implement MPC.

To account for multiple factors in model design~(objective, RNN/Transformer/CNN, uni/bidirectional), we consider the implementations of APC, MPC, and CPC as listed in Table~\ref{tab:model_notations}.

\subsection{Pre-training datasets}
We use LibriSpeech~\cite{panayotov2015librispeech}, which contains about~1k hours of speech audio, for pre-training all self-supervised models.
We also use the {\tt \small unlab-6k} subset from Libri-Light~\cite{kahn2020libri}, which contains about~6k hours of speech audio, for additional experiments in Section~\ref{sec:effect}.
We use~80-dimensional log Mel spectrograms as input acoustic features, i.e.,~$\mathbf{x}_{t}\in \mathbb{R}^{80}$.
All models are trained for~10 epochs using Adam with a batch size of~32 and a learning rate of~$10^{-3}$.
During pre-training, only the speech portion from the dataset is used.

\subsection{Probing datasets}
\myparagraph{Representation similarity measures}
For calculating representation similarity with {\tt \small lincka} and {\tt \small svcca}~(introduced in Section~\ref{sec:sim}), we use the {\tt \small si284} subset from Wall Street Journal~(WSJ)and the train set from TIMIT.

\myparagraph{Phonetic and speaker classification}
We carry out both classification tasks on WSJ.
For phonetic classification, there are a total of~42 phone categories, and we follow the standard split of WSJ, using~90\% of {\tt \small si284} for training,~10\% for validation, and reporting frame-level phone error rate on {\tt \small dev93}.
The phone alignments are generated with a speaker adapted GMM-HMM model.
For speaker classification, we follow~\cite{chung2020generative} and consider a~259-class classification task where each class corresponds to an unique speaker, using~80\% of {\tt \small si284} for training, the other~10\% for validation, and reporting utterance-level speaker error rate on the rest~10\%.
We note that speaker classification is not a typical task for WSJ, and only serves as a sanity check for the presence of speaker information.
For both tasks, the classifier is a linear logistic regression trained for~10 epochs using SGD with a batch size of~32 and a fixed learning rate of~$10^{-4}$.
All reported error rates are an average of~5 runs, of which variances are negligibly small and not included.

\section{Results and Analysis}
\label{sec:results}

\subsection{Similarity of different self-supervised representations}
\label{sec:sub_similarity}
\begin{figure}[htbp]
  \centering
  \includegraphics[width=\columnwidth]{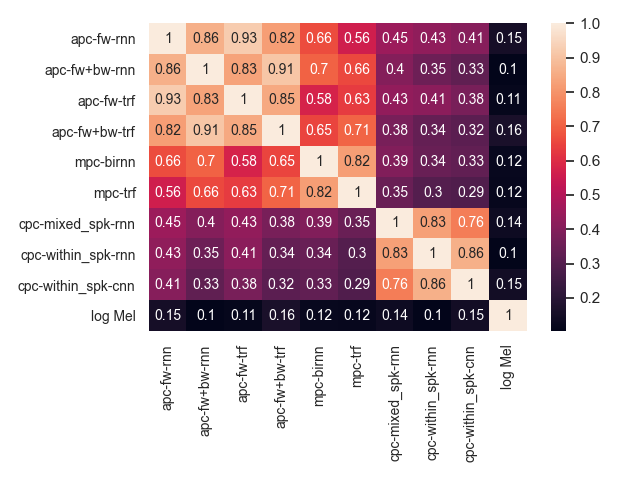}
  \caption{Similarity heatmap of various self-supervised representations on WSJ according to {\tt \small lincka}. Similarity values are also annotated.}
  \label{fig:wsj_similarity}
\end{figure}

Figure~\ref{fig:wsj_similarity} shows the heatmap of similarities between representations learned by various self-supervised models. Brighter colors indicate higher similarity between two representations.
We also include the similarity between each self-supervised representation and the surface feature, i.e., log Mel spectrogram.
Due to space limit, we only show the result according to similarity measure {\tt \small lincka} on WSJ.%
\footnote{More similarity heatmaps are available at \url{https://github.com/iamyuanchung/ICASSP21-Similarity-Supplementary}}
We find all heatmaps exhibiting consistent patterns regardless of the probing dataset and similarity measure, and all self-supervised representations are very different from the surface feature.
The heatmaps reveal the following insights.

\myparagraph{Objective affects similarity more than architecture.}
The most evident pattern from the heatmaps is that there is always a greater similarity within an objective than across objectives, indicated by the bright block diagonal.
For example, {\tt \small apc-fw-rnn} is always more similar to {\tt \small apc-fw+bw-rnn}, {\tt \small apc-fw-trf}, and {\tt \small apc-fw+bw-trf} than to any MPC and CPC variants, even when {\tt \small apc-fw-rnn} and {\tt \small cpc-mixed\_spk-rnn}/{\tt \small cpc-within\_spk-rnn} share the same building block and directionality.
This conclusion also holds for the MPC- and CPC-family.
Representations learned by generative-based objectives, i.e., variants of APC and MPC, are also more similar to one another than to the CPC variants.

\myparagraph{Directionality affects similarity more than building block.}
When the objective is the same, we find that model's directionality~(uni/bidirectional) has a higher impact on representation similarity than its building block~(RNN/Transformer/CNN).
For instance, the similarity between {\tt \small apc-fw-rnn} and {\tt \small apc-fw-trf}, which are both unidirectional while the former uses RNNs and the latter uses Transformers, is higher than that between {\tt \small apc-fw-rnn} and {\tt \small apc-fw+bw-rnn}, which both use RNNs while the former is unidirectional and the latter is bidirectional.
Furthermore, as may be expected, making APC bidirectional reduces its difference with MPC, which is indicated by the fact that {\tt \small mpc-birnn} is more similar to {\tt \small apc-fw+bw-rnn} than to {\tt \small apc-fw-rnn}, and {\tt \small mpc-trf} is more similar to {\tt \small apc-fw+bw-trf} than to {\tt \small apc-fw-trf}.

\myparagraph{Source of negative samples affects similarity more than architecture.}
When focusing on the CPC-family, we find that the proposal distribution of where the negative samples are drawn from is more impactful on representation similarity than the building block.
This is indicated by the fact that {\tt \small cpc-within\_spk-rnn} is more similar to {\tt \small cpc-within\_spk-cnn} than to {\tt \small cpc-mixed\_spk-rnn}, where the two models in the former case share the same proposal distribution but use different building blocks, and the two models in the latter case share the same building block but incorporate different proposal distributions.

\subsection{Correlation between self-supervised loss and phonetic \& speaker classification performance}
\label{sec:corr}
Experiments so far have only revealed the similarities between different self-supervised representations.
We further uncover the correlation between self-supervised loss during pre-training and the amount of phonetic and speaker information contained in the representations, measured by their performance on phonetic and speaker classification (Section~\ref{sec:probe}).
\begin{table}[htbp]
  \caption{Pearson correlation coefficients between the self-supervised loss and the phone and speaker error rates.~$*$ denotes statistical significance at~$\rho < 0.05$.}
  \label{tab:corr}
  \centering
  \begin{tabular}{lcc}
    \toprule
    Model                             &     Phone     &    Speaker   \\
    \midrule
    {\tt \small apc-fw-rnn}           &  0.989$^{*}$  &  0.950$^{*}$ \\
    {\tt \small mpc-birnn}            &  0.885$^{*}$  &  0.847$^{*}$ \\
    {\tt \small cpc-mixed\_spk-rnn}   &  0.643$^{*}$  &  0.762$^{*}$ \\
    {\tt \small cpc-within\_spk-rnn}  &  0.675$^{*}$  &  -0.071      \\
    \bottomrule
  \end{tabular}
\end{table}
We only consider {\tt \small apc-fw-rnn}, {\tt \small mpc-birnn}, {\tt \small cpc-mixed\_spk-rnn}, and {\tt \small cpc-within\_spk-rnn} in this experiment for a comparison only in terms of their objectives~(except {\tt \small mpc-birnn}, which has to be bidirectional).
We calculate the Pearson correlation coefficients~$r$ between loss value and both phone and speaker error rates, as listed in Table~\ref{tab:corr}.

\vfill\pagebreak

Overall, APC and MPC are found to have a stronger correlation between the self-supervised loss and both their phonetic and speaker classification performance than CPC.
In particular, {\tt \small apc-fw-rnn} features the strongest correlation among the four considered self-supervised models.
Our finding aligns with~\cite{blandon2020analysis}, where the autoregressive loss of APC is found to be more correlated with the ABX-score of a phone discrimination task than the InfoNCE loss of CPC.

It is noteworthy that the loss of {\tt \small cpc-within\_spk-rnn} has almost no correlation with speaker classification performance.
This result seems natural since the model always draws negative samples from the same utterance as the positive sample, so speaker information is never found to be useful for distinguishing them and thus not learned by the representation.
On the other hand, the proposal distribution of {\tt \small cpc-mixed\_spk-rnn} allows the model to learn from negative samples coming from both the same and different utterances as the positive sample, meaning that both phonetic and speaker information could be relevant for discriminating them.
Therefore, we find the loss of {\tt \small cpc-mixed\_spk-rnn} is still correlated with the speaker error rate to some degree.

We emphasize that our findings here are not meant to claim any self-supervised approach to be the best, but aim to provide some results for other researchers for future reference.
For example, APC and MPC's strong correlation between their self-supervised loss and phonetic and speaker classification performance could be useful for model selection even during the pre-training stage, since a lower pre-training loss would indicate a richer phonetic and speaker representation.
CPC, though exhibiting a smaller correlation between its self-supervised loss and phonetic and speaker classification performance, could still be extremely powerful when the downstream task is known and thus the pre-training proposal distribution can be determined aforehand, as shown by its recent impressive performance on semi-supervised speech recognition~\cite{baevski2020wav2vec}.

\subsection{Effect of increasing unlabeled data for pre-training}
\label{sec:effect}
One of the biggest advantages of self-supervision is its capability to leverage large-scale unlabeled data for representation learning.
Here we train {\tt \small apc-fw-rnn}, {\tt \small mpc-birnn}, {\tt \small cpc-mixed\_spk-rnn}, and {\tt \small cpc-within\_spk-rnn} on~2k,~4k, and~6k hours of speech audio, all sampled from the {\tt \small unlab-6k} subset of the Libri-Light corpus, and calculate the similarities between each of these variants and their counterpart trained on the original~960 hours LibriSpeech audio according to {\tt \small lincka}.
Results are shown in Table~\ref{tab:more_data_sim}.

\begin{table}[htbp]
  \caption{Representation similarity between self-supervised models pre-trained on $\sim$1k hours of audio and their counterparts pre-trained on increasing amounts of audio according to {\tt \small lincka}.}
  \label{tab:more_data_sim}
  \centering
  \begin{tabular}{lccc}
    \toprule
    \multirow{2}{*}{Model}            &  \multicolumn{3}{c}{Hours of pre-training audio} \\
    \cmidrule(lr){2-4}
                                      &  $\sim$2k  &  $\sim$4k  &  $\sim$6k \\
    \midrule
    {\tt \small apc-fw-rnn}           &  0.957  &  0.935  &  0.923 \\
    {\tt \small mpc-birnn}            &  0.940  &  0.939  &  0.925 \\
    {\tt \small cpc-mixed\_spk-rnn}   &  0.911  &  0.883  &  0.837 \\
    {\tt \small cpc-within\_spk-rnn}  &  0.920  &  0.896  &  0.861 \\
    \bottomrule
  \end{tabular}
\end{table}

For all models, their representations become more dissimilar when more data are used for pre-training.
We also find that CPC's representations change more than those of APC and MPC when increasing the data size.
For instance, the similarity ``only'' drops from~0.957 to~0.923 for {\tt \small apc-fw-rnn} when increasing the data size from~2k hours to~6k hours, while for {\tt \small cpc-mixed\_spk-rnn}, the similarity drops from~0.911 to~0.837.

Changes in representation similarity can be attributed to encoding details of speech other than phonetic and speaker information that might be unnecessary, such as background noises.
To confirm whether such changes in representation similarity correspond to an actual richer phonetic and speaker representation, we again use phonetic and speaker classification performance to quantify the amount of phonetic and speaker information contained in the representation.

Encouragingly, we observe that most self-supervised models' performance on both tasks is improved when being trained on more data.
The only exception is {\tt \small cpc-within\_spk-rnn} on speaker classification, which is expected as speaker information is never found relevant for discriminating positive and negative samples during its training.
However, its performance on phonetic classification obtains the largest gain among all considered self-supervised models.
Concerning {\tt \small cpc-mixed\_spk-rnn}, in addition to showing improvement on both tasks, the drop of its speaker error rate
is also the largest among all models.
Intuitively, having more data means that CPC models are provided with more comparisons of negative and positive samples to learn from, and our results seem to suggest that this is a more effective way for learning representations when large amounts of unlabeled data are available, as opposed to attempting to reconstruct details of the speech signals as APC and MPC models do.
That being said, both generative- and contrastive-based objectives also benefit from having more unlabeled training data.

\section{Conclusions}
\label{sec:cons}
We have analyzed representations learned by contrastive predictive coding~(CPC), autoregressive predictive coding~(APC), and masked predictive coding~(MPC) through the lens of similarity analysis.
Extensive experiments have been conducted to study the impact of different modeling choices for training self-supervised models, the effect of the size of unlabeled training data, and how well the self-supervised loss correlates with phonetic and speaker classification performance.
We have found that the self-supervised objective has a much higher impact on representation similarity than architectural choices such as building blocks~(RNN/Transformer/CNN) and directionality~(uni/bidirectional).
We have also observed that APC has the strongest correlation between its self-supervised loss and phonetic and speaker classification performance, which is useful for model selection.
Finally, while all self-supervised models benefit from having more training data, CPC is found to learn from the additional data more efficiently than APC and MPC.

\section{Acknowledgments}
Yonatan Belinkov was supported in part by the ISRAEL SCIENCE FOUNDATION (grant No. 448/20) and by an Azrieli Foundation Early Career Faculty Fellowship. 

\vfill\pagebreak

\bibliographystyle{IEEEbib}
\bibliography{strings,refs}

\end{document}